\documentclass[11pt,twoside]{article}
\usepackage{./asp2014}
\usepackage{graphicx}

\aspSuppressVolSlug
\resetcounters

\begin{document}

\title{An analysis of the $V$-band light curve of the Be star $\omega$ CMa with the viscous decretion disk model}
\author{M.R. Ghoreyshi,$^{1\star}$ A.C. Carciofi$^1$
\affil{$^1$Instituto de Astronomia, Geof\'{\i}sica e Ci\^{e}ncias Atmosf\'{e}ricas, Universidade de S\~{a}o Paulo, Rua do Mat\~{a}o 1226, Cidade Universit\'{a}ria, S\~{a}o Paulo, SP 05508-900, Brazil; $^{\star}$\email{mohammad@usp.br}}}

\paperauthor{M.R. Ghoreyshi}{mohammad@usp.br}{}{Instituto de Astronomia, Geof\'{\i}sica e Ci\^{e}ncias Atmosf\'{e}ricas}{Universidade de S\~{a}o Paulo}{S\~{a}o Paulo}{SP}{05508-900}{Brazil}

\begin{abstract}
We analyze the $V$-band photometry data of the Be star $\omega$ CMa, observed during the last four decades. The data is fitted by hydrodynamic models based on the viscous decretion disk (VDD) theory, in which a disk around a fast-spinning Be star is formed by material ejected by the central star and driven to progressively wider orbits by means of viscous torques. For the first time, we apply the model for both the disk build up and the dissipation phases. Our simulations offer a good description of the photometric variability in both phases, which suggests that the VDD model adequately describes the disk structural evolution. Furthermore, our analysis allowed us to determine the viscosity parameter ($\alpha$) of the gas, as well as the net mass loss rate. We find that $\alpha$ is variable, ranging from 0.1 to 1.0, and that build up phases have larger values of $\alpha$ than the dissipation phases. Additionally, we find that, contrary to what is generally assumed, even during quiescence the outward mass flux is never zero, suggesting that the star alternates between a high mass-loss phase (outburst) and a low mass-loss phase (quiescence).
\end{abstract}

\section{Introduction}\label{sec1}


\begin{figure}[!h]
\centering
\includegraphics[width=1.0\linewidth]{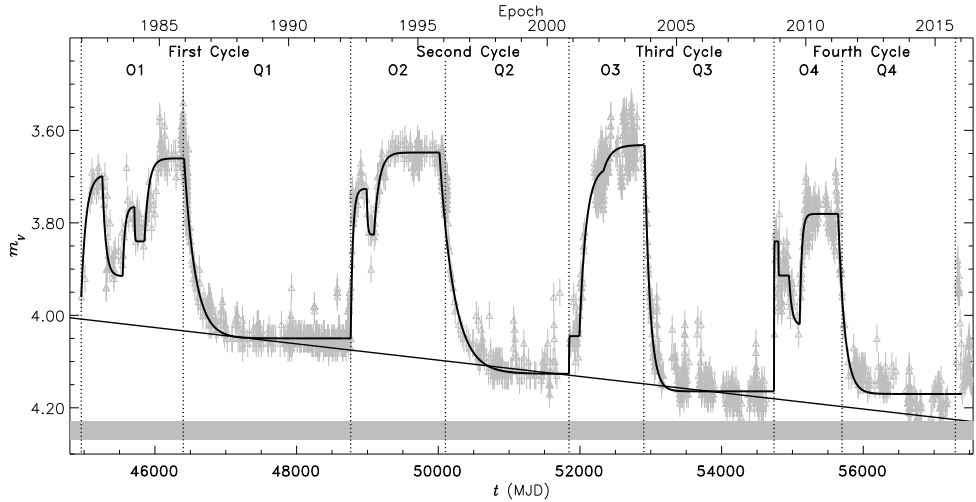}
\caption{$V$-band light curve of $\omega$ CMa (grey triangles), showing four full cycles as indicated, alongside by the exponential fit presented in Sect. \ref{sec2.1} (solid line). The slant line shows a linear fit of the last 100 days of each cycle, indicating the long-term secular decline in the brightness of the system in successive dissipation phases. O$i$ and Q$i$ indicate outburst and quiescence, respectively, where $i$ is the cycle number. The light curve is a collection of observations from the following sources: \citet{baade1982}, \citet{stagg1987}, \citet{dachs1988}, \citet{edalati1989}, photoelectric observations obtained in the LTPV (\citealt{manfroid1991}, \citeyear{manfroid1995}; \citealt{sterken1993}), \citet{mennickent1994}, Hipparcos project (\citealt{perryman1997}), \citet{stefl2000}, and the visual observations by Otero \citep{stefl2003}. The horizontal grey band represents the intrinsic visual magnitude of the $\omega$ CMa, as estimated using the stellar parameters of Table \ref{table:1}. The vertical dotted lines indicate the transitions between quiescence an outburst (and vice-versa).}
\label{fig1}
\end{figure}

Be stars are massive, fast rotating stars with dust-free Keplerian gaseous disks around them. Because there are numerous Be stars in the solar neighborhood (there are $\sim$ 100 listed as such in the Bright Star Catalogue), there are a large number of Be stars with a rich observational history. As a result, their circumstellar disks have been observed by a variety of different techniques.

Many Be stars were observed in phases of disk construction and dissipation, during which the brightness of the star varies as a result of the changing amount of circumstellar material. However, a quasi cyclic photometric behavior is a characteristic which can be seen in the light curve of only a few of them. $\omega$ CMa ($m_{v}$ = 3.6 to 4.2, Fig. \ref{fig1}) is one such star that shows an alternating 2.5 $\sim$ 4.0 years of outburst (during which a new disk forms and the star gets brighter in the visible) and 4.5 $\sim$ 6.5 years of quiescence (as the disk dissipates and the star dims). $\omega$ CMa is a B3Ve star that has been a common target of observers due to its peculiarities. Therefore, there exists a rich data-set that has been investigated by several authors. Consequently, the stellar parameters are very well known (Table \ref{table:1}).

The disk around Be stars are more likely described by the same mechanism found in the accretion disks of young stellar objects, namely viscous transport of material and angular momentum. However, while in accretion disks the direction of flow is inwards, in Be stars it can alternate between outflow (decretion, when mass is actively being lost by the central star) and inflow (when the disk reaccretes back to the star when it ceases to lose mass). One open issue of such systems is the origin of viscosity: so far no satisfactory theory of viscosity exists and for this reason the $\alpha$ prescription for viscosity is commonly assumed \citep{shakura1973}. The $\alpha$ parameter links the scale of the turbulence to the (vertical) scale of the disk by the following formula:

\begin{equation}
\label{eq1}
\nu = \alpha c_{s}H,
\end{equation}
where $c_{s}$ is the sound speed and $H$ is the disk scale height. Measuring the value of $\alpha$ as direct as possible will help astronomers to have a better understanding about the physics of viscosity and also of viscous disks, and this is the main motivation for this work.

Using the VDD model \citep{lee1991}, which has become the current paradigm to explain the disks around Be stars \citep{rivinius2013}, \citet{carciofi2012} modeled the dissipation phase of $\omega$ CMa's light curve from 2003 to 2008, and derived for the first time the viscosity parameter, $\alpha$, of a Be star disk.

\begin{table}[!h]
\begin{center}
\hspace*{-1. cm}
\begin{tabular}{@{}cccccccccccc}
\hline
\hline
Parameter & \vline & $L$ & $T_\mathrm{pole}$ & $R_\mathrm{pole}$ & log g$_\mathrm{pole}$ & $M$ & $V_\mathrm{rot}$ & $V_\mathrm{crit}$ & $R_\mathrm{eq}$ & $i$ & $m^{*}_{v}$ \\
 & \vline & (L$_{\odot}$) & (K) & (R$_{\odot}$) & & (M$_{\odot}$) & (km s$^{−1}$) & (km s$^{−1}$) & (R$_{\odot}$) & & \\
\hline
& \vline \\
Value & \vline & 5224 & 22000 & 6.0 & 3.84 & 9.0 & 350 & 436 & 7.5 & 15$^{\circ}$ & 4.25 \\
\hline
\end{tabular}
\caption{The stellar parameters of $\omega$ CMa \citep{maintz2003}.}
\label{table:1}
\end{center}
\end{table}

In the work of \citet{carciofi2012}, $\alpha$ was determined only during a phase of disk dissipation. To advance upon the work of Carciofi et al. we modeled the full light curve of $\omega$ CMa since 1982 to 2015 including several outburst and quiescence phases. We begin analysing the lightcurve with an exponential formula, in order to extract some parameters such as the durations of each phase, without engaging in the difficulty of a physical modeling (Sect. \ref{sec2.1}). We then proceed to describe the results of a full VDD modeling in Sect. \ref{sec2.2}.


\section{Models and Results}\label{sec2}

\subsection{Exponential Fitting}\label{sec2.1}

We fitted the $V$-band data of $\omega$ CMa by using the following exponential formula:

\begin{equation}
\label{eq3}
m_\mathrm{v}(t) = (m_{0}-m^{\prime})e^{-(t-t_{0})\tau^{-1}}+m^{\prime},
\end{equation}
where $m_\mathrm{v}(t)$ is the visual magnitude as a function of time, $m_{0}$ is the visual magnitude at the beginning of each section ($t=t_0$), $m^{\prime}$ is the asymptotic visual magnitude of each section, and $\tau$ is the associated timescale of the brightness variation.

Based on the exponential formula fitting (Fig. \ref{fig1}), the intersection point of subsequent curves provides one way of estimating the starting and ending time of each event and thus estimate the total length of each event. This information will be used as initial parameters for the VDD modeling. The results show that the length of successive formation phases and successive dissipation phases are decreasing.

We found that the growth and decay timescale of the light curve, $\tau$, varies from cycle to cycle and also within a given cycle. This result suggests that the viscosity parameter, $\alpha$, is varying, as $\alpha$ is the main parameter controlling the disk evolution timescales, the other parameters being the stellar mass and the disk temperature. In the next section we compare the values of $\tau^{-1}$ to the ones of $\alpha$ obtained by the VDD fitting of the data, and indeed find a tight correlation between them.

\subsection{Viscous Decretion Disk Model}\label{sec2.2}

For the physical modeling of the light curve of $\omega$ CMa we used one of the most performing radiative transfer codes to date, {\tt HDUST} \citep{carciofi2006a}, with the time-dependent hydrodynamics code, {\tt SINGLEBE} \citep{okazaki2007} that provides the disk surface density and the mass flow rate as a function of radius at each selected time, which will be used as input for the {\tt HDUST} code to calculate the spectral energy distribution and then the $V$-band excess of the disk as a function of time. More details on {\tt SINGLEBE} and on the adopted boundary conditions can be found in \citet{haubois2012}.

\begin{figure*}[!h]
\begin{minipage}{1.0\linewidth}
\centering
{\includegraphics[width=1.0\linewidth, height=0.4\linewidth]{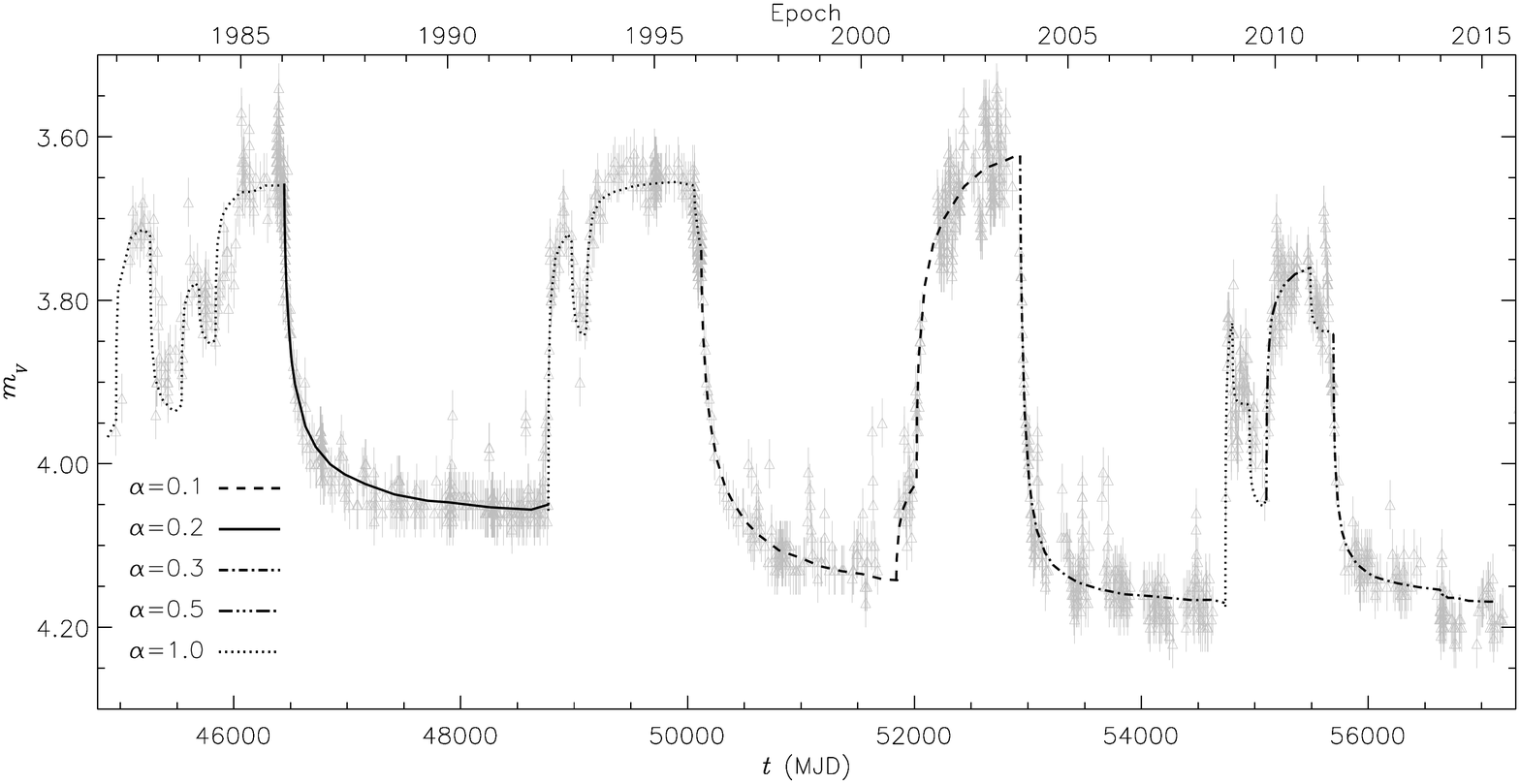}}
\end{minipage}\par\medskip
\begin{minipage}{1.0\linewidth}
\centering
{\includegraphics[width=1.01\linewidth, height=0.4\linewidth]{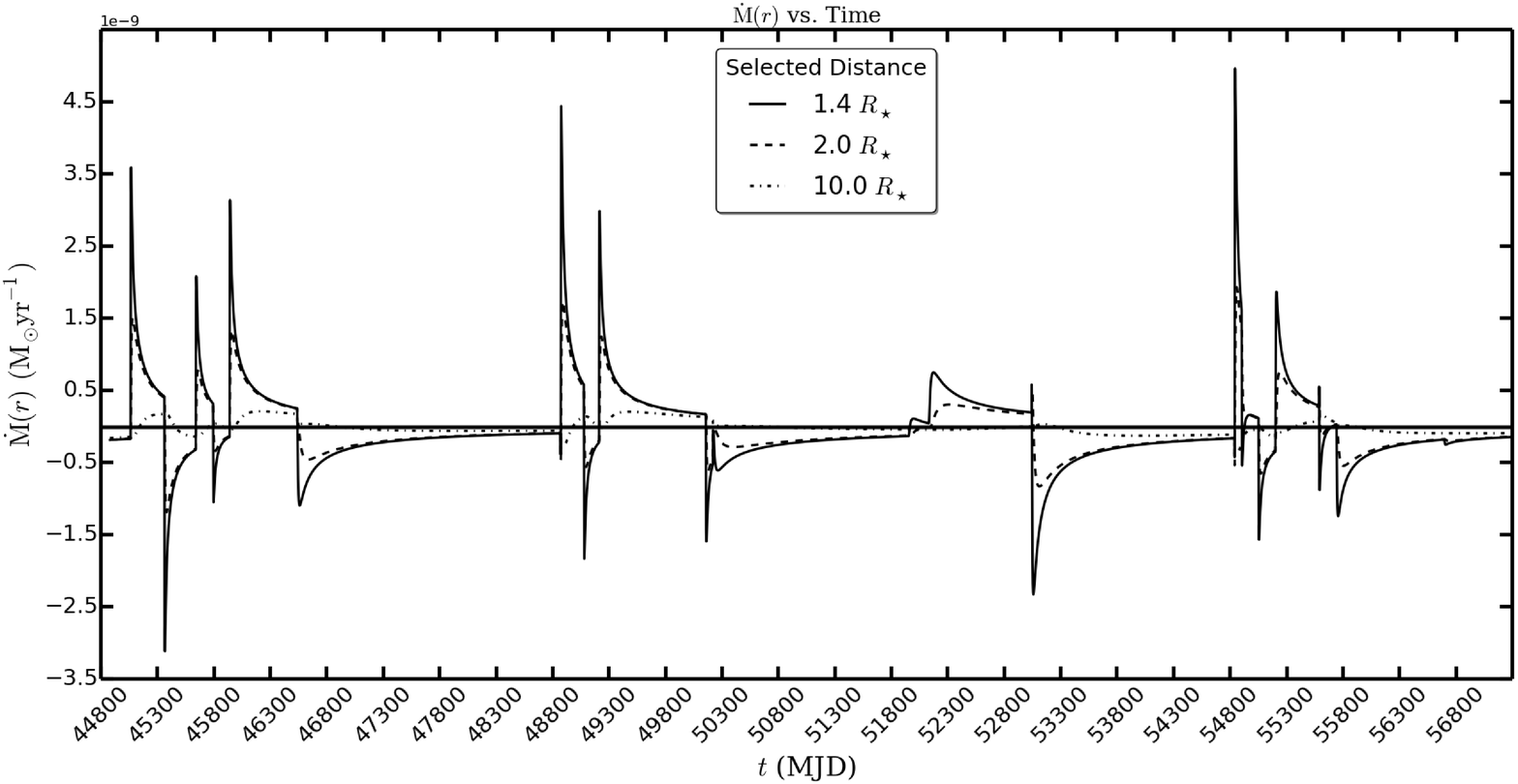}}
\end{minipage}\par\medskip
\begin{minipage}{1.0\linewidth}
\centering
{\includegraphics[width=1.0\linewidth, height=0.4\linewidth]{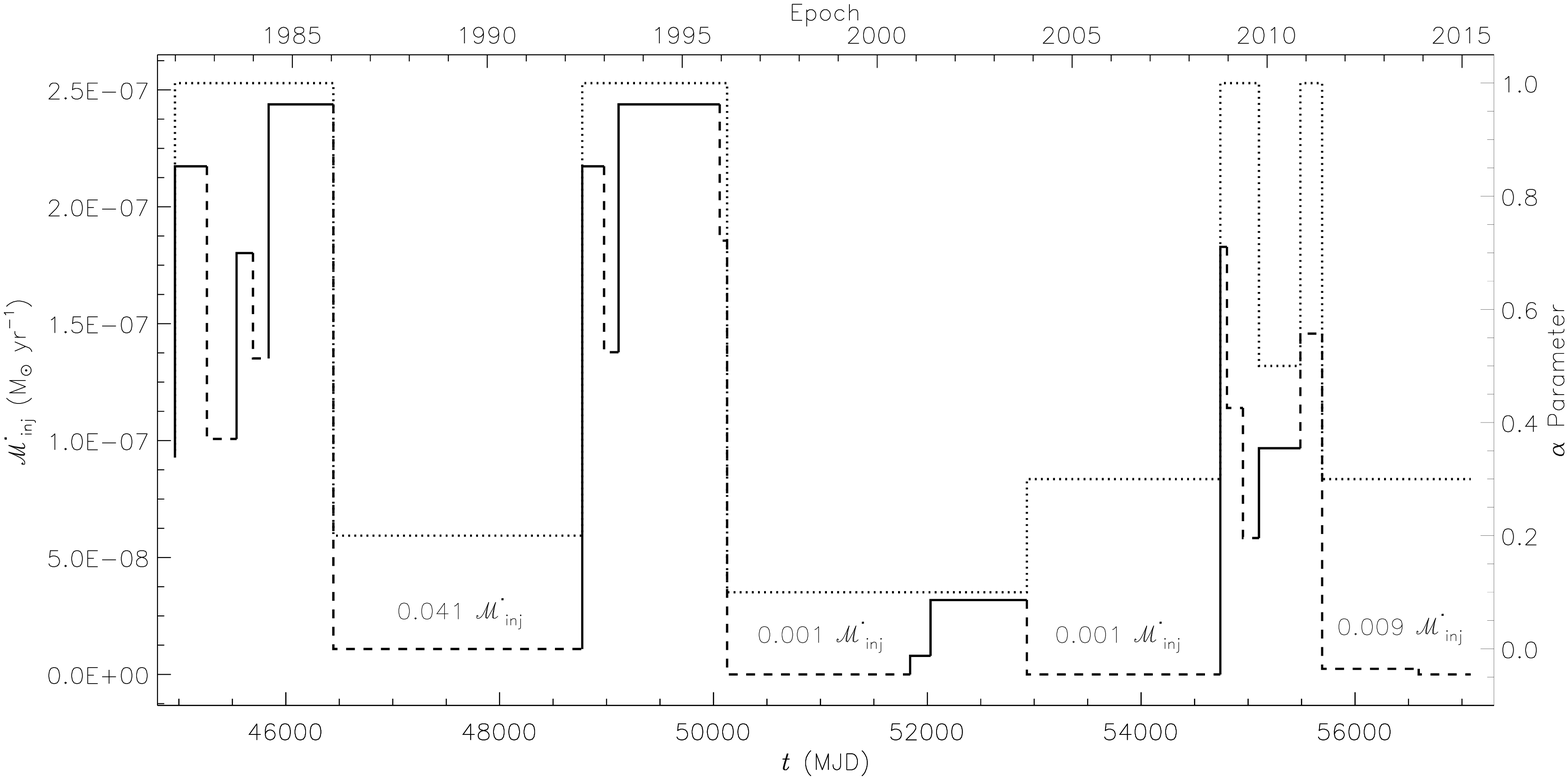}}
\end{minipage}
\caption{
{\it Top}: VDD model of the full light curve of $\omega$ CMa. Each line type represents an individual value for $\alpha$ parameter, as indicated.
{\it Middle}: Mass flux as a function of time, at different positions in the disk. The solid, dashed, and dash-dotted lines corresponds to $r = 1.4 \mathrm{R}_{\star}$, $r = 2.0 \mathrm{R}_{\star}$, $r = 10.0 \mathrm{R}_{\star}$, respectively.
{\it Bottom}: The mass injection rate history of the best fit model. The solid lines show the outbursts and the dashed lines show quiescence. The dotted lines represent the value of $\alpha$.}
\label{fig2}
\end{figure*}

For each part of the light curve, as defined in the previous section, the free parameters are the mass injection rate, the $\alpha$ parameter, and the onset time of each phase, and these three are explored in order to find the best match between model and observation. Finally, we consider that at quiescence the star shifts to a lower mass injection rate instead of entering a state of absolute quiescence. With this scenario we obtained a good fit for the light curve of $\omega$ CMa (Fig. \ref{fig2}). We found that $\alpha$ is variable in the range of 0.1 to 1.0, and it is larger at build up phases than in dissipation ones. Moreover, the mass injection rate varies between 3.2 $\times$ 10$^{-08}$ M$_{\odot} \mathrm{yr}^{-1}$ to 2.4 $\times$ 10$^{-7}$ M$_{\odot} \mathrm{yr}^{-1}$ in outburst and 2.7 $\times$ 10$^{-10}$ M$_{\odot} \mathrm{yr}^{-1}$ to 1.1 $\times$ 10$^{-08}$ M$_{\odot} \mathrm{yr}^{-1}$ in quiescence.

Fig. \ref{fig2} illustrates an important property of the VDD model. The bottom panel shows $M_\mathrm{inj}$, which is the mass injection rate at the base of the disk. The middle panel shows the mass flux as a function of time at different positions of the disk. The mass flux is always much smaller than the mass injection rate (in the order of 100 times), which means that most of the ejected matter falls back onto the star.
Fig. \ref{fig2} also shows that the disk changes between decretion/accretion phases, where $\dot{M}(r)$ is positive/negative.

\begin{figure}
\centering
\includegraphics[width=1.0\linewidth]{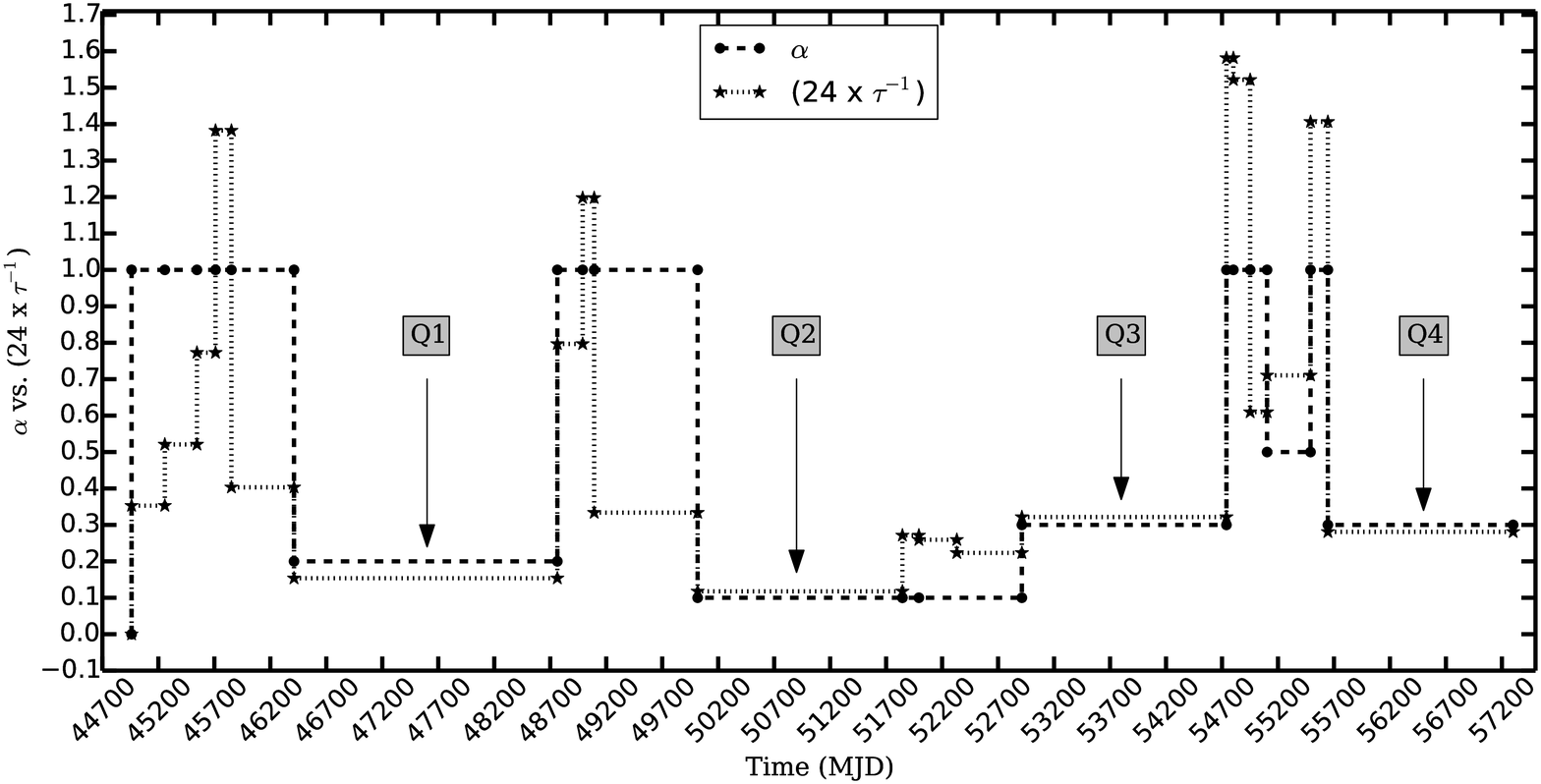}
\caption{The comparison between the values derived for $\alpha$ (dashed lines) and $\tau^{-1}$ (dotted lines) parameters.}
\label{fig3}
\end{figure}

Fig. \ref{fig3} compares the values of $\tau^{-1}$ (dotted lines) and $\alpha$ (dashed lines). There is a correlation between $\alpha$ and $\tau^{-1}$, as expected from the fact that $\alpha$ is the main parameter controlling the disk timescales. Fig. \ref{fig3} suggests an exponential fit as a possible way to obtain the $\alpha$ parameter, provided that the $\tau^{-1}$ parameter can be calibrated from hydrodynamical calculations. This will be the subject of a future study \citep{rimulo2017}.

\section{Conclusions}\label{sec3}

We modeled the $V$-band light curve of $\omega$ CMa by two different models: mathematical (exponential) and physical (VDD) models. The main result of this work is that it demonstrates, for the first time, that the VDD model is capable of reproducing the disk variability both at build up and dissipation phases. Another quite intriguing result is that different values of $\alpha$, in the range of 0.1 $\sim$ 1.0, were required at different phases with higher values at disk formation phases.

Moreover, our model was able to reproduce the decline in the brightness of the system in successive dissipation phases (slant line in Fig. \ref{fig1}). We suggest that one way to interpret this behavior is to consider that at quiescence the star shifts to a lower disk-feeding rate instead of entering a state of absolute quiescence (i.e., zero disk-feeding rate).

\end{document}